
\input phyzzx

 \def\frac#1#2{{\textstyle{#1\over #2}}}

 \def\FigCap{\par\penalty-100\medskip \spacecheck\sectionminspace
    \line{\fourteenrm\hfil FIGURE CAPTIONS\hfil}\nobreak\vskip\headskip }

\pubnum={NI-94014}
\titlepage
\title{A modified discrete sine-Gordon Model }
\author{ Wojtek J. Zakrzewski\foot{email:w.j.zakrzewski@durham.ac.uk}\break}
\address{Isaac Newton Institute\break
20 Clarkson Road\break
Cambridge, CB3 0EH, England\break}
\centerline{\it and}
\DURHAM

\abstract
We modify the recently proposed model of Speight and Ward to make it
possess time dependent solutions.  We find that for each lattice spacing
and for each velocity of the sine Gordon kink we can find a modification
of the model for which this kink is a solution. We find that
this model has really 3 ``kink-like" solutions; the original
kink, the static kink and a further kink moving with
velocity $v\sim0.97$.
We discuss various properties of the model, from the point of
view of its usefulness for numerical simulations.
\chapter{Introduction}
Recently Speight and Ward,
in a very interesting paper,\Ref\rmartin{J.M. Speight and R.S. Ward - {\it
Nonlinearity} {\bf 7}, 475-484 (1994)}
 presented a discrete version of the Sine-Gordon model which had
topologically stable kink-like solutions. Their idea was to introduce
 a discrete version of the Bogomolnyi bound\Ref\rBogo{E.B. Bogomol'nyi - {\it
Sov. J. Nucl. Phys.} {\bf 24}, 449 (1976)}
and then use it for deriving a bound on the energy of a
kink in the model.
The field configuration which saturates this bound has
the lowest energy and is topologically stable.

Their work starts from the observation that in the continuum
$${1\over 4}\int_{-\infty}\sp{\infty}(\psi_x\sp2+sin\sp2\psi)\;dx
={1\over 4}\int_{-\infty}\sp{\infty}(\psi_x-sin\psi)\sp2\;dx
+{1\over 2}\int_{-\infty}\sp{\infty}(\psi_xsin\psi)\;dx\,\eqn\eenergycon$$
where $\psi_x={\partial \psi\over \partial x}$, $\psi={1\over 2}\phi$ and
where $\phi$ is the  familiar field of the sine-Gordon model\Ref\rsine{see \eg\
P.G. Drazin and R.S. Johnson - Solitons: an Introduction - C.U.P. (1989)}
\ie\ the model whose time-dependent equation of motion is
$$\phi_{tt}-\phi_{xx}=sin\phi.\eqn\esinegordon$$
 Hence as $$\int_{-\infty}\sp{\infty}(\psi_xsin\psi)\;dx=
\int_{-\infty}\sp{\infty}-(cos\psi)_x\;dx=cos(\psi(-\infty))-cos(\psi(+\infty))
\eqn\eboundcont$$
we see that in the class of field configurations which have ``kink-like"
boundary conditions, \ie, which satisfy $\psi(-\infty)=0,\;
\psi(+\infty)=\pi$ the lowest energy field must solve
$$\psi_x=sin\psi.\eqn\ebogcont$$

When the continuum is replaced by a spatial lattice we have to decide
how to replace the derivatives $\psi_x$ \etc; the most naive idea
would correspond to using the forward
differences \ie\ puting $\psi_x={\psi(x+h)-\psi(x)\over h}$, where
$h$ is the lattice spacing; however, this choice is very nonunique and it
is this nonuniqueness that was exploited by Speight and Ward.
Namely, they observed that  the key line in the argument given above
corresponds to
$${\partial (cos\psi)\over \partial x}\,=\,-sin\psi\,{\partial \psi\over
\partial x}.\eqn\ecos$$
They noted that if the model is discretised in such a way that
the factorisation implied by \ecos\ holds and ${\partial cos\psi\over
\partial x}$ is replaced by the finite difference of $cos\psi$
the discretised model will possess the topologically stable
kinks. As $$cos(\psi(y))-cos(\psi(x))=
-2sin\bigl({\psi(y)-\psi(x)\over 2}\bigr)
sin\bigl({\psi(y)+\psi(x)\over 2}\bigr),
\eqn\efact$$
this has led them to replace
$$sin(\psi(x))\rightarrow\,
sin\bigl({{\psi(x+h)+\psi(x)}\over 2}\bigr)\eqn\esin$$
and $${\partial \psi\over
\partial x}
\rightarrow\,{2\over h}sin\bigl({{\psi(x+h)-\psi(x)}\over 2}\bigr).\eqn\eder$$
Of course, both terms have the correct $h\rightarrow 0$ limit.

With this replacement the model possesses stable kink like solutions.
They are given by the solutions of the discrete analogue of \ebogcont;
namely, $$sin\Bigl({{\psi(x+h)+\psi(x)}\over 2}\Bigr)=
{2\over h}sin\Bigl({{\psi(x+h)-\psi(x)}\over 2}\Bigr).\eqn\ebogdiscr$$

In fact,
Speight and Ward
have managed to find the analytical form
of this discrete kink-like solution
which allowed them to
discuss its various properties. In particular, they pointed out that
as the solution possesses a translational zero mode
(despite the lack of translational symmetry); this
mode can then be used to study the dynamics
of the kinks in the collective coordinate approximation.

To do this they had to introduce the time dependence into the
model; for that they added the usual term \ie\ $\bigl(
{\partial \psi\over \partial
t}\bigr)\sp2$ to the Lagrangian density and then calculated
the corresponding equations of motion. In their case
this equation took the form of
$$\psi_{tt}={1\over h\sp2}\bigl[sin(\psi_+-\psi)-sin(\psi-\psi_-)\bigr]
-{1\over 4}\bigl[sin(\psi_++\psi)+sin(\psi+\psi_-)\bigr],\eqn\eequation$$
where $\psi=\psi(x),\;\psi_{\pm}=\psi(x\pm h)$.

Speight and Ward studied the dynamics of their kinks using this equation.
The time evolution was simulated by the fourth order Kutta-Runge method.
Moreover, the results were compared with the results obtained
in the collective variable approximation.

First of all, the results showed that the kink in the model
behaved very much like the kink in the continuum; the
Peierls Nabarro barrier of the usual discrete models\Ref\rpeierls{see \eg\
J.F. Currie, S.E.  Trullinger, A.R. Bishop and J.A. Krumhansl  - {\it Phys.
Rev.}
{\bf B 15}, 5567 (1977) or R. Boesch, C.R. Willis and M. El-Batanouny - {\it
 Phys. Rev.}
{\bf B 40}, 2284 (1989)}
 was eliminated
by the novel discretisation and hence the model was a much
better approximation of the model in the continuum.
Moreover, the results showed that the collective approximation worked very well
for small velocities and for small lattice sizes (\ie\ for small $h$).
For larger velocities the kink radiated and slowed down. The results
were more pronounced for larger values of $h$; in fact there was a
natural lattice-size ``cut-off" ($h=2$) above which the whole model
was unstable. Although the ``cut-off" value of $h$ was ridiculously
high (and so irrelevant from the point of view of any ``physical"
applications) its existence can be traced to Speight and Ward's choice
of their discretisation \esin\ and \eder. However, their choice was
one of many (albeit perhaps the most ``natural" one); other choices
change this value and can even eliminate it altogether.

In this paper we discuss these choices and point out that the kinetic
term can also be altered. We then present a model which
not only possesses static kinks but also
moving kinks of the continuum sine-Gordon type
moving at one velocity which is related to the choice
of the parameters of the model. We discuss their
dynamics and find that the model preserves all the good points
of the original model of Speight and Ward, has considerably reduced
radiation effects and so could be used for the numerical study of the dynamics
of kinks in the continuum had we not been in the lucky situation of having
analytical solutions or for other numerical investigations.

\chapter{Modified Models}
\section{Static case}
The choice of Speight and Ward \esin\ and \eder\ is clearly very nonunique.
The idea of using \efact\ to factorise it into the right hand terms
of \esin\ and \eder\ can clearly be modified; \eg\ by the inclusion
an arbitrary factor $f(x,h)$ in one term and its inverse in the other;
\ie\ we can put
$$sin(\psi(x))\rightarrow\,
f\;sin\bigl({{\psi(x+h)+\psi(x)}\over 2}\bigr)\eqn\ensin$$
and $${\partial \psi\over
\partial x}
\rightarrow\,{1\over f}\,{2\over h}sin\bigl({{\psi(x+h)-\psi(x)}\over
2}\bigr).\eqn\ender$$
What can $f(x,h)$ be? Clearly, we need $lim_{h\rightarrow 0}f=1$
but otherwise we have complete freedom. On the other hand, any
explicit $x$ dependence in $f$ would be somewhat undesirable since it would
break the translational symmetry of the lattice.

The Speight and Ward choice corresponds to $f=1$ but we not have to
be so drastic. Of course, any choice of $f$ will have implications
for the static kink solution as the Bogomolnyi bound equation
now becomes
$$sin\Bigl({{\psi(x+h)+\psi(x)}\over 2}\Bigr)f\sp2=
{2\over h}sin\Bigl({{\psi(x+h)-\psi(x)}\over 2}\Bigr).\eqn\ebogdiscrnew$$
If we now take
$$f\sp{2}={h\over 2}{(1+e\sp{h})\over (e\sp{h}-1)}\eqn\efspeight$$
then, as is easy to check, \ebogdiscrnew\ has
$$\psi(x)=2\,arc\,tan\{exp(x-a)\},\eqn\ekinkusual$$
\ie\ the usual kink, as its solution.
Here, as in the continuum, $a$ is arbitrary
and can be thought of the ``position"
of the kink (which, in general, falls
between the lattice sites). Hence, like in the case of Speight and Ward,
the kink solution possesses a continuous translational zero mode.

Of course \efspeight\ has a correct
$h\rightarrow 0$ limit, is monotonic and for large $h$ $f\sim \sqrt{h\over 2}$.
Hence, as we compare the models with different lattice spacing $h$
we see that the effect of $f(h)$ is to decrease the importance of the
$sin(\psi)\sp2$ term in the energy density at large $h$.
Of course, the total energy of \ekinkusual\ is  still the same as
in the continuum, \ie\ 1 in our units. The kink is stable, and any
other field configuration (with the same boundary conditions)
has larger energy.

What about the dynamics of the kink? Will it propagate
and will it radiate? The work of Speight and Ward has partially answered
this question. Using \eequation\ they studied the dynamics of their
kink and found that it radiated (and the amount of radiation
depended on the velocity and on the lattice spacing).
However, the choice of the time dependent term in \eequation, although
``natural" is again very nonunique; one can consider taking
other terms except that in the limit $h\rightarrow 0$ we should
recover $\psi_{tt}$. Any such modification will have implication
on the kinetic term in the Lagrangian density and so also on the
total energy of a given field configuration.

So what can we choose? Clearly, we should not change the degree
of the equation of motion; so all modifications should involve
only the additional dependence on $\psi_t$, $\psi$ and $h$. As there
is very little guidance as to what to take we have looked
first at the case of Speight and Ward and then considered
modifications involving the multiplications of $\psi_{tt}$
by a function $g(\psi_t,\psi,h)$.
However, the simplest modifications would involve $g$ independent of $\psi$,
as then we do not modify too dramatically the Lagrangian density  - so, in this
paper we restrict ourselves to considering only $g=g(\psi_t,h)$.

\section{Time Dependent Case}
The simplest modification, which is not too drastic would involve
the replacement
$$\psi_{tt}\,\rightarrow\;{\psi_{tt}\over (1+\alpha\psi_t\sp2)},\eqn\enewa$$
where $\alpha$ is a function of $h$, such that
$\alpha(0)=0$. Such a modification leads
to the Lagrangian density which consists of two terms (the kinetic energy
term (which involves only ${\partial \psi(x,t)\over \partial t}$)
 and the potential term
which involves $\psi(x,t)$, $\psi(x+h,t)$ and $\psi(x-h,t)$.

Hence, the equation of motion which we want to study, and which is
a generalisation of \eequation\ is

$${\psi_{tt}
\over (1+\alpha\psi_t\sp2)}={f\sp2\over
h\sp2}\bigl[sin(\psi_+-\psi)-sin(\psi-\psi_-)\bigr]
-{1\over
4f\sp2}\bigl[sin(\psi_++\psi)+sin(\psi+\psi_-)\bigr],\eqn\enewequation$$
where $\alpha$ and $f$ are functions of $h$, such that
$\alpha(0)=0$ and $f(0)=1$ and, as before, $\psi_{\pm}=\psi(x\pm h)$.

As it is easy to check \enewequation\ is the equation of motion
which can be derived  from the following form of
the Lagrangian  $L=L_k-L_p$, where
$$L_k=h\sum_i\Bigl\{
{\psi_t(i)\over 2\sqrt{\alpha}}\,arctan(\sqrt{\alpha}\psi_t(i))
-{1\over 4\alpha}ln(1+\alpha\psi_t(i)\sp2)\Bigr\}\eqn\ekinetic$$
and
$$L_p=h\sum_i\Big\{{f\sp2\over h\sp2}sin\sp2\bigl(
{\psi(i+1)-\psi(i)\over 2}\bigr)
+{1\over 4f\sp2}sin\sp2\bigl({\psi(i+1)+\psi(i)\over 2}\bigr)
\Bigr\},\eqn\epotential$$
 where we have now put $\psi(i)=\psi(x,t)$,
$\psi(i\pm 1)=\psi(x\pm h,t)$ and
 $\psi_t(i)={\partial \psi(x,t)\over \partial t}$.

Our expression for $L_k$ has a very unusual look but, as can be checked, is
definitive positive. As $arctan(x)\sim x-{x\sp3\over 3}+..$
and $ln(1+x)\sim x-{1\over 2}x\sp2+..$ for small $x$
we see that for small  values of $\sqrt{\alpha}\psi_t$
$${\psi_t(i)\over 2\sqrt{\alpha}}\,arctan(\sqrt{\alpha}\psi_t(i))
-{1\over 4\alpha}ln(1+\alpha\psi_t\sp2(i))
\sim \psi_t\sp2(i)-{1\over 2}\psi_t\sp2(i)
={1\over 2}\psi_t\sp2(i)\eqn\ekineticsmall$$
and so we recover the conventional behaviour.
The differences do show  up as $\sqrt{\alpha}\psi_t$
increases. In fig 1. we present plots of the
two expressions of $L_k$ (\ie\ the summand in
\ekinetic, and the right hand side of
\ekineticsmall) as a function of $\psi_t$ (for $\alpha=3.7768$).
We see that for $\psi_t$ up to about 0.2 the two expressions agree
and for larger values of $\psi_t$ \ekinetic\ grows with $\psi_t$
less fast than \ekineticsmall. In fact it is easy to see that
for larger values of $\psi_t$ the first term in \ekinetic\ dominates
more and more and so, for large values of $\psi_t$ the dependence
on $\psi_t$ becomes essentially linear. However, as we will discuss
in the next section, in practical applications $\psi(t)$ never gets very
large and so our values of $L_k$ will not be
that different from their usual values.

In the discussion above we have not identified $L_k$ with
the kinetic energy of the system. One may be tempted to do so, but
this would be incorrect as $L_k$ is not quadratic in $\psi_t$.
To find the correct expression we need to calculate the Hamiltonian
and then take its part which involves $\psi_t$. Then we find that
the kinetic energy $E_{kin}$ is given by
$$E_{kin}=h\sum_i\Bigl(\psi_t(i){\partial L_k\over \partial \psi_t(i)} -
L_k\Bigr)
=h\sum_i\Bigl({1\over
4\alpha}ln(1+\alpha\psi_t(i)\sp2)\Bigr),\eqn\erealkinetic$$  \ie\ by only the
$log$ term in \ekinetic\ taken with the opposite sign.
Hence in fig.1 we also plot the $log$ term. We see that its growth with
$\psi_t$ is even slower than the other two terms.

\chapter{A Propagating Kink}
Can we find any solution of \enewequation? Of course this depends
on our choices of the functions $f(h)$ and $\alpha(h)$.

However, let us observe that the usual (\ie\ continuum) expression
for the Sine-Gordon kink
$$\psi(x,t)=2arctan\bigl\{exp\bigl({x-vt-x_0\over \sqrt{1-v\sp2}}\bigr)\bigr\}
\eqn\eusualkink$$
has a chance of being a solution  as the time derivatives of \eusualkink\
and the terms on the right hand side of \enewequation\ tend
to give relatively similar expressions.
In fact, a couple of pages of algebra shows that \eusualkink\
is a solution of \enewequation\ if
$\alpha$ is chosen to be given by
$$\alpha(h)={(1-v\sp2)\over v\sp2}sinh\sp2(2\beta)
\eqn\ealphavalue$$
and $f(h)$ is given by
$$f\sp{2}={{v\sp2\over 1-v\sp2}+\sqrt{\bigl({v\sp2\over 1-v\sp2}\bigr)\sp2
+{4sinh\sp2(2\beta)\over h\sp2}}\over
{8sinh\sp2\beta\over h\sp2}},\eqn\ef$$
where $\beta={h\over2\sqrt{1-v\sp2}}$.
Thus we see that we can have a moving kink but
its velocity is fixed by the model;
namely is determined by $\alpha$ and $f$.
Note that our expressions for $\alpha$  and $f$ have
the correct $h\rightarrow0$
limit ($\alpha\rightarrow 0$ and $f\rightarrow 1$). On the other hand
if we keep $h$ fixed and take $v\rightarrow0$
$$f\sp2\rightarrow {h\over 2}coth\bigl(h),
\eqn\elimf$$
\ie\ we obtain the expression we have mentioned before \efspeight.
At the same time, however, $\alpha\rightarrow\infty$; so we cannot use
\ef\ for $v$ too small. In practice, this does not matter as we will
argue below, and we will use $v=0.7$.

What is the behaviour of $f\sp2$ as we vary $h$ and $v$?
It is easy to check that for small $h$ and for small $v$ $f\sp2\sim1$.
If we keep $v$ not too large (say below (0.8)) $f\sp2$ grows
with $h$ but this growth is very small until $h\sim 2$.
For larger values of $v$ (say above 0.85) $f\sp2$ is close to 1
for small $h$ then decreases and reaches a minimum around $h=1.5$, and then
becomes very similar to $f\sp2$  for smaller $v$.
This is presented in fig. 2 where we have plotted the dependence
of $f\sp2$ on $v$ for some selected values of $h$.
Hence,  for almost any value of $h$ the dependence on $v$,
as long as $v<0.9$, is very small
and so we may expect that motions with different values
of velocity will not be qualitatively very different.
We discuss this in the next section.

However, before we do that, let us observe that our model, in addition
to  having moving a kink-like solution, possesses also  a static solution
which satisfies the Bogomolnyi bound.

Static solutions are solutions of \ebogdiscrnew\ for $f\sp2$ given by \ef.
They can be easily found by observing that  \ebogdiscrnew\
is equivalent to
$$tan\bigl[{\psi(x+h)\over 2}\bigr]=tan\bigl[{\psi(x)\over 2}\bigr]
{c+1\over c-1},\eqn\eneww$$
where $c={2f\sp2\over h}$. Hence puting $\psi(x)=2arctan\{u(x)\}$
we see that \eneww\ reduces to
$$u(x+h)=u(x){c+1\over c-1},\eqn\enewww$$
which has as its solution $u(x)=\alpha exp\{\beta (x-x_0)\}$
where $\beta$ is given by
$$\beta={1\over h}\,ln\,\bigl\{{1+c\over c-1}\bigr\},\eqn\enewwww$$
and where $x_0$ is arbitrary.
Hence, the static solution is again in the form of a moving
kink with its profile determined by $\beta$. To see the
values of $\beta$ \enewwww\ gives let us put $h=1$ and $v=0.7$ in \ef\
(as these are the values we will use in the next sections).
Then we find $\beta=1.02257$, which in turn corresponds to $v=0.208938$.
Hence a static kink, in our model with $h=1$ and $v=0.7$ in \ef,
is given by a kink whose profile has been deformed as if it were
moving with velocity $v=0.208938$. It is interesting that this velocity
is so low. It is easy to check that the potential energy of such a static
kink is indeed 1 (in fact, numerically, we find 1.0000000004).

\chapter{Numerical Simulations}
Before we present and discuss our results let us point out that
we have really two ``obvious" modifications of the original model
of Speight and Ward. The first one, for the static case,
is the one  in which
we can take $f\sp2={h\over 2}coth\bigl(h)$ and then the
time dependent one with $f\sp2$ given by  \ef. Of course, we
could also consider fixing $f\sp2$ at its value given by \ef\
and then study kinks whose profile is given by $\beta$ close
to $\beta\sim 1.02257$ (\ie\ corresponding to the static kink
mentioned above) and which move at some other (small) velocity.
However, we have not studied such motions in much detail as we
expect them to be not very dissimilar from the case when
$f\sp2={h\over 2}coth\bigl(h)$.

Hence, restricting our attention first
to the case when $f\sp2={h\over 2}coth\bigl(h)$
we see that this modification can be used for the study of static
or slowly moving kinks in the cases when the lattice spacing $h$
has to be large. In the static case we have the topological bound, \ie\
the potential energy is always bounded from below by 1 and
equal to 1 for the static field which solves the Bogomolnyi bound
\ebogdiscrnew. When we consider the slowly moving kinks, and introduce
the motion by a nonrelativistic boost the potential still satisfies
the Bogomolnyi bound but now the field starts radiating readjusting
itself to its new shape. The readjustment is related to the velocity
of the kink; it grows with the velocity.
This is the radiation observed by Speight and Ward.

On the other hand we can start the kink with a modified profile
(by a Lorentz factor); then the kink needs less readjustment
and so it radiates less. However, as soon as we start considering
moving kinks it makes more sense to use $f\sp2$ given by \ef.
In this case we can perform the simulation with the initial velocity
of the kink being close to the value of the velocity used in the
definition of $f\sp2$ or being quite  different. We would expect our results
to depend on the value of this velocity; the outgoing radiation
is expected to increase with the mismatch of velocities.
Moreover, as soon as we use $f\sp2$ with $v\ne0$ the field
configuration does not saturate \ebogdiscrnew\ and so the
potential energy is different from 1.

Also, even when we take the same value of the velocity as in the definition
of $f\sp2$ we expect some radiation; this is due to the fact
that our kink solves \enewequation\ in which time is continuous
while the numerical simulation (fourth order Kutta Runge) uses
a finite (albeit very small) $dt$.

Next we present some results of our simulations. Most of the simulations
refer to the case when $f\sp2$ was given by \ef\ with the value of $v$
 taken as $v=v_1=0.7$. Later we looked at other values too
but as there was very little difference all
 our results  in this paper refer to this value of $v_1$.

We have performed many numerical simulations (on grids of different
sizes) varying both $h$, $v$ (of the actual kink) and using
both expressions for the kinetic energy (conventional and \erealkinetic.)
In all simulations involving
one kink we used \eusualkink\ to calculate from it $\psi(0)$ and
$\psi_t(0)$ and then used these quantities as our initial value
data.

Of course, for small values of $h$ all the results
of our simulations were essentially
indistinguishable from each other (for the same
value of $v$); hence let us concentrate our discussion
on the case when $h$ was sizeable, say $h=1$.

First we look at the motion of the kink for which $v=v_1$.
In fig. 3a we present plots of the field at $t$=0,
400, 800, 1200, 1600 and 2000.
To put all the fields into one picture the fields
at $t\ne0$ have been displaced by subtracting $0.2*{t\over 400}$
from their values. Our results were obtained
on the grid stretching from -2000 to +2000; the time step of our simulation
was 0.001.
In fig. 3b we present the plot of our field at $t=1200$ for $800<x<880$
as it is only in this region that the field is substantially
different from its asymptotic values (0, and $\pi$). Note that our
kink is described by essentially 6-10 lattice points, the others
are very little different from their asymptotic values.
 We see essentially no radiation.

 However, although we see virtually no radiation, strictly speaking,
the radiation is nonzero
(due to  the $t$ discretisation);  this is exhibited
in fig. 3c where we plot  the field at $t=800$
for $-600<x<-100$. Note the vertical scale which
demonstrates the smallness
of the radiation.

 We also looked at the velocity
with which the kink was propagating along the lattice. Of course, as the kink
is described by a few points on the lattice it is difficult to decide what
is meant by its velocity and how to calculate it.
 We  have used two methods to calculate it. The first method involved
 finding
the best approximate position of the kink
$x(t)$ at each value of $t$ (by a linear interpolation between
the largest value of $x$ for which the field was still negative and
the smallest $x$ for which it was positive); the velocity  was then
approximated by the rate of change of this position function $x(t)$.
As the extrapolation is not very exact
the resultant velocity has a tendency to oscillate  a little and so, to be more
precise,
we should average the obtained results.

 The other method
involved calculating the ``average relevant" $X(t)$
by defining
$$X(t)={\sum_i x(i)\,eng(i,t)\over \sum_i eng(i,t)}\eqn\eaveragex$$
where $eng(i,t)$ was the total energy of the lattice point $i$.
Then we could use the rate of change of this $X(t)$ to calculate the velocity.
This  approach does not involve any extrapolation and is more ``natural", but
it involves not only the kink but the whole field so
it takes into account also radiation effects.
Thus as long as only few lattice points are involved
in the dynamics the two expressions
give the same value for the velocity and the approximation based on $X(t)$
is smoother; when the radiation effects become appreciable $X(t)$
is pulled down by the radiation waves moving to the left and the resultant
velocity does not measure the velocity of the kink but a kind of
``overall" velocity of the field.

In fig. 4 we present the time dependence of the velocity
 calculated by the second
method (the result of the first method is essentially indistinguishable
except for some initial oscillations). We see that,
as expected, the kink propagates
with $v=0.7$ with no slowing down.

Next we looked at the kinks started  of with different values of $v$; \ie\
we used
\eusualkink\ with different values of $v$ to calculate our initial data.
The results can be summarised as follows; for most velocities below
$v=v_1$ there is very little radiation; the kink propagates with a velocity
a little lower than $v$ and radiating slows down. The rate of the slowing down
is very low.
For $v>v_1$ the kink starts propagating with velocity larger than $v$,
also radiates and slows down. This time the radiation effects are more
pronounced but still very small; hence even now the slowing down
is not very pronounced. Thus we see that we have a model in which kinks
radiate little.
We may wonder why the initial velocity of the kink is not $v$ (except
for $v=v_1$). This is due to the readjustment of the kink; the initial
configuration does not fit with the lattice distribution and so the kink
readjusts itself and travels with velocity which corresponds to the
speed of this readjusted field. Of course we do not expect these effects to
be too large and this is what we have observed in our simulations.

In fig. 5 we present plots of the time dependence of the velocity of the
kinks initiated with different values of $v$. We see that all kinks
have radiated very little, in agreement with our expectations. This should
be contrasted with the results obtained by Speight and Ward. They
only studied kinks moving at small velocities, but even then their
kinks radiated more. The results in fig. 5 present the results
 for $t$ up to  200;
one may wonder what happens when the kinks propagate longer. Speight
and Ward, in their paper, present the  time dependence of the
velocity of a kink initially moving at $v=0.28$. Their plot shows
relatively rapid slowing down which later decreases slightly. To see what
happens in our case we rerun some of our simulations for much longer times.
To avoid any boundary condition problems we extended the lattice sufficiently
far; this has extended the time needed for the completion of each
simulation. In fig. 6a we present the plots of the fields
at $t=0$, 800, 1600, 2400, 3200 and 4000 of the kink
started off with $v=0.25$. As before they have been
successively displaced down by 0.2.
In fig. 6b we present the plot of the time dependence of the kink's velocity
seen in this simulation.
 The plot shows two curves corresponding to our two
ways of evaluating the velocity of the kink. As expected, there is very
little difference between the two curves; the slowing down is not
very substantial over the whole period of our simulation. Of course, there
is some radiation and so the curve of the time dependence of the velocity
evaluated with $X(t)$ lies lower. However, the two curves
 differ very little; the
fact that there is little radiation is confirmed by looking at fig. 6c
where we exibit the details of the field at $t=4000$ for $900<x<970$.

Our results should be contrasted with the results obtained using our
lattice and the standard  expression for the kinetic energy and
the model of Speight and Ward.
In fig. 6d we present the time dependence of the velocity of the kink  as seen
in all these types of simulations. We find that both other schemes
lead to much more radiation and so in both of them the kink slows down
much more.

Looking at the velocities shown in fig. 5 we see that the
simulation started at $v=0.8$ involved more radiation; hence we looked
at it in detail. In fig. 7 we present the fields at 4 values of time
($t=0$, 800, 1600 and 2400) for this case
(again the fields are displaced vertically). We see that
this time  the radiation
is more visible.

For the simulations with $v$ around $0.9$ we find, once again, very little
slowing down of the kink, hence in fig. 8a we present  the fields
at some values  of time for the simulation with $v=0.9$. In fig. 8b
we present the plot of the time dependence of the kink. We see
that the radiation effects are very small and that the kink
propagates with velocity $v\sim0.978$.

For larger values of $v$ (say $v=0.95$ or larger) we again have bigger
radiation effects - fig. 9a shows
the fields at 3 values of time
for $v=0.95$. For $v=0.97$, given the coarseness
of the effective lattice (so that initially effectively
only 3 lattice  points are involved), the radiation
effects are so large that the kink becomes unstable and it develops
a jump of $\pi$
at $x\sim 0$. Given the coarseness of the lattice this
is easy to achieve but the upshot of this is that, unexpectedly,
 the kink begins
to move in the opposite direction.
In fig. 9b we present the fields at 3 values of time (0, 80 and 160).
One can calculate the velocity of the kink and  find that it is about -0.3.
Hence we see that for large velocities and for coarse lattice cases
the system becomes unstable; not surprisingly, as initially only few
lattice points are involved, they can move up or down by $\pi$ and
so the dynamics of the kink could become very different from what may be
expected from its continuum configuration.

\chapter{Other field configurations}
Given our lattice we can check whether it allows the propagation
of other kink-like configurations. In particular, had we not known
the analytic form of the
two-kink and other field configurations we could  have used it for such
 numerical studies. To answer this question, at least partially, we
decided to look at the field configurations describing two moving
 kinks  and a breather.
\section{Two kinks}
First we looked at the configuration involving two kinks.
As is well known\refmark\rsine the analytic form of such a solution is given by
$$\psi(x,t)=2\, arctan\Bigl\{{vsinh{x\over \sqrt{1-v\sp2}}\over
cosh{vt\over \sqrt{1-v\sp2}}}\Bigr\},\eqn\etwokinks$$
where $\pm v$ is the velocity of each kink relative to $x=0$.

We have looked at the time evolution of such field configurations;
as before, we used \etwokinks\ to calculate the initial condition
$\psi(0)$ and $\psi_t(0)$ which were then used in our simulations.
In fig. 10 we plot the field configurations at some chosen values
of time obtained in a simulation with $v=0.3$.
We note some
radiation, but again not very much of it; in fig. 11 we plot the velocity
of the kinks which confirms that the kinks do not slow down too much.
We have performed simulations with kinks started off
at other velocities ($v$=0.1,
and 0.9); the
resultant simulations were very similar in each case; in each simulation
the kinks first accelerated to reach their maximum speed (which was
always a little lower than $v$), then slowed down a little, although
this slowing down was hardly perceptible. The two effects, the slowing
down and not reaching $v$ were partly due to the radiation.
\section{breather}
We have also looked at the behaviour of the breather in our model.
The initial condition was taken from the analytic form of the breather,
namely\refmark\rsine it was given by
$$\psi(x,t)=2arctan\Bigl[{\sqrt{1-\lambda\sp2}\over \lambda}
{sin\{\lambda(t-t_0)\}\over
cosh\{\sqrt{1-\lambda\sp2}x\}}\Bigr],\eqn\ebreather$$
where $\lambda$ describes the frequency of the breather. $t_0$ was chosen
to be ${\pi\over 2\lambda}$ so that at $t=0$ the field was temporarily at rest.
Several simulations were performed with different values of $\lambda$.
This time, by comparison with the  case of one and two kinks,
the radiation
effects were more important but still
not that important.
The amount of the radiation depended on the period
of the breather, which is determined by $\lambda$.
 For small values
of $\lambda$ (say $\lambda=0.25$)
the breather radiated with the amplitude
of the field at the origin gradually decreasing. However, this decrese
was very pronounced only at the very begining
of the simulation but later became hardly
noticeable. In fig. 12a we plot the time dependence of the value
of the field at $x=0$ as seen in our simulation. In fig. 12b we present
our fields at 6 values of time (0, 400, 800, 1200, 1600 and 2000)
successively displaced in the vertical direction down by 2 units.
We see that the breather radiates, and this radiation then propagates
out. Gradually the breather radiates less and less and this is consistent
with the slow decrease of its amplitude. In fig. 12c we replot 12b
by restricting $x$ to $-50<x<50$. This shows the decrease of radiation
in the immediate neibourhood of the breather itself.

In fig. 13 we present  similar results for $\lambda=0.9$. This time
the radiation effects are much smaller; the amplitude hardly
decreases and we see very little radiation  in the fields
themselves (this time displaced up). In fig. 13c, this time, we show
the details of the field at $t=2000$ for this case from which we see
that, although weakly, the breather radiates all the time.

 Similar results were seen in other simulations.

\chapter{Some general comments and conclusions}
The model we have used in our study is based on the model of Speight and Ward.
It retains many nice features of their model and, when used in
simulations on very coarse lattices, it leads to considerably reduced
radiation effects.

In fact, studying various properties of the model in  a one kink sector
we see that the model has really three stable  ``kink-like"
solutions; a moving kink
(in our case moving at $v=0.7$), a static kink (the usual kink whose
profile has been modified so that it looks as a kink moving with
$v\sim 0.209$) and a kink moving with $v\sim0.97$.

Thus, as such, the model may be  used as a tool for studying
the interactions of kinks. In particular, it may be used to study
the interactions of fast moving kinks, which in the more
conventional models are plagued with big radiation effects.
In our simulations we have set $v=v_1=0.7$. There was nothing special
about this value; we have also performed simulations at other values
of $v_1$; the results were similar; although too low a value  of $v_1$
(like $v_1\sim0.1$) led to increased radiation effects due to the higher
value of $\alpha$ and hence increased nonlinearity in $\psi_t$.
For $v\sim v_1$ the radiation effects were  negligible and the kink
propagated along the lattice with virtually no slowing down.

The price we have paid for having a  discrete model
with moving kinks was, of course,  the somewhat
unusual form of the kinetic energy of the kink;  however, its
properties were not that different from the usual energy and the
distinction was noticeable only at higher speeds.

When we compare  energies of the
kinks which were started off
at different speeds  the effects due to the unconventional
form of the kinetic energy became significant for larger velocities.
In fig 14 we present
the plots of the total energy in simulations for different values
of $v$ (the two curves represent the usual expression for the
kinetic energy, and ours). As they include the energy of the radiation
they cannot be thought of as the energy of the kink, but at least
initially, they are close to it. We see the same trend of both
expressions; however, the energy of our kink grows with the increase
of velocity slower than in the conventional case.

One of the main advantages of the model of Speight and Ward was that
the fields  in this model satisfied the Bogomolnyi bound. In our case,
 because of the
explicit dependence on $v$ in the expression for $f$, this was  no
longer true;
however, in practice, the potential energy, given by $L_p$ in (2.9),
was almost constant in time
and varied very little with $v$. This is shown in fig. 15. As before the curves
have been vertically displaced by 0.2 units upwards.
We see very little variation  of $L_p$ with time. This is, of course,
in agreement with our expectations. The exact moving kink solution
describes a kink which displaces itself along the lattice. The kink
is not on the flat bottom of Bogomolnyi solutions and so the orbit
of the kink is not an equipotential curve, but instead, the potential
varies periodically along this orbit.
The variation is very small indeed.
It is reassuring to see that
the total energy was  extremely well conserved
(to within 9 decimal points); hence the kinetic energy
varied periodically too, in precise antiphase with the potential.
 In addition $L_k$, the time dependent term in the
Lagrangian,  also varied periodically with $t$, but its amplitude
was  greater.

Hence we see that we have a model which can be used for numerical
investigations of kink-like structures and their
 interactions on coarse lattices.
As such it reproduces the continuum behaviour very well; although at very large
velocity and for very coarse lattices the radiation effects can become
significant and can alter the behaviour of the kinks.
The model is also reasonably successful in reproducing the behaviour
of periodic structures, like breathers.

At the same time, the model may become interesting in itself for studying
the dynamics of discrete systems. In this case  we have to decide whether
we should take seriously our choice of the kinetic term.
The term is unusual but, as such,  it does not violate any
basic principles. However, its unusual form suggests caution. Thus if
we cannot justify its use then, perhaps,  we should  return to
the static form of $f$ and the
conventional form of the kinetic term. However, then we would expect
the radiation effects to become more pronounced. But whether this
is physically justified or not can only be determined by
looking in detail at the
concrete applications.

\ack
\par
 A large part of the work reported in this paper
 was performed when WJZ was visiting
 the Centre de Recherches Mathematiques at the Universit\'e de Montreal
in Canada.
 He wishes to thank Pawel Winternitz, and other members of the Centre for their
 hospitality, support and for their interest.

The work was finished at the Isaac Newton Institute in Cambridge. The author
would like to thank the Institute for its kind hospitality and  for provinding
a stimulating research environment.

He also wishes to thank Martin Speight and Pawel Winternitz
 for useful conversations.

Finally, he wishes to acknowledge
a grant from the Nuffield Foundation which has made
his visit to Montreal possible.
\FigCap
\par
{ \par \parindent= 0pt
\item{Fig 1:} Comparison of our time dependent term in the Lagrangian ($L_k$)
{\it (a)}
\ekinetic\ with the conventional
energy \ekineticsmall\  {\it (b)} and with the kinetic energy ($E_k$) {\it (c)}
for different values of $\psi_t$
\item{Fig 2:}  $v$ dependence of $f\sp2$ given by \ef\ for $h=0.1$, 0.3, 0.5,
0.7, 1, 2, 3 and 4. For small $v$ $f\sp2$ monotonically increases with $h$.
\item{Fig 3:} Fields  as seen in the simulation with $v=0.7$, a) Fields
at $t=0$, 400, 800, 1200, 1600 and 2000 displaced successively  down by 0.2.
 b) Field
at $t=1200$ for $800<x<880$, c) Field at $t=800$ for $-600<x<-100$.
\item{Fig 4:} Time dependence of the velocity of the kink as seen
in the simulation of Fig. 3.
\item{Fig 5:} Time dependence of the velocity of the kinks as seen
in the simulations initiated with $v$=0.01, 0.1, 0.2, 0.3, 0.4, 0.5, 0.6, 0.7,
0.8, 0.85 and 0.9.
\item{Fig 6:} Simulation initiated with $v=0.25$. a) Fields at $t=0$, 800,
1600, 2400, 3200 and 4000 displaced successively  down by 0.2.
b) Time dependence of the velocity
of the kink; two method of evaluation $\alpha$) by extrapolation
$\beta$) in terms of $X(t)$, c) Field at $t=4000$
for $900<x<970$, d) Time dependence of the velocity of the kink as seen in our
simulation ($\alpha$), simulation in our model but with the conventional
expression for the kinetic energy ($\beta$), and the model of Speight and Ward
($\gamma$).
\item{Fig 7:} Fields, at $t=0$, 800, 1600 and 2400 as
seen in the simulation with $v=0.8$. The fields are successively
displaced down by $0.4$.
\item{Fig 8:} Fields (a), at $t=0$, 400, 800,  1200, 1600 and 2000 as
seen in the simulation with $v=0.9$. The fields are successively
displaced down by $0.2$. b) Time dependence of the velocity of the kink.
\item{Fig 9:} Fields at $t=0$, 80 and 160 as seen in simulations with
larger $v$. a) $v=0.95$, b) $v=0.97$. Again, fields have been displaced
vertically.
\item{Fig 10:} Simulations of two kinks. Fields at $t=0$, 400, 800 and 1200
successively displaced up by 0.4 for a simulation initiated with $v=0.3$.
\item{Fig 11:} Time dependence of each kink as seen in the simulation shown
in Fig 10.
\item{Fig 12:} Simulation of the breather with $\lambda=0.25$. a) Time
dependence of the field at $x=0$, b) Fields at $t=0$, 400, 800, 1200,
1600 and 2000, successively displaced down by 2. c) As b) but restricted
to $-50<x<50$.
\item{Fig 13:} Simulation of the breather with $\lambda=0.9$. a) Time
dependence of the field at $x=0$, b) Fields at $t=0$, 400, 800, 1200,
1600 and 2000, successively displaced up by 2. c) Detail of the field
at $t=2000$ .
\item{Fig 14:} Comparison of the velocity dependence of
the energies of one kink; a) with the conventional term, b) with our term
\ie\ \ekinetic.
\item{Fig 15:} Time dependence of potential energies as seen in our
simulations of one kink initiated with $v=0.1*i$ for $i=1,..9$
again displaced successively up by 0.2.

}
\refout

\end
\bye